\documentclass[apjl]{emulateapj}
\def\ltsima{$\; \buildrel < \over \sim\;$}
\def\ltsim{\lower.5ex\hbox{\ltsima}}
\def\gtsima{$\; \buildrel > \over\sim \;$}
\def\gtsim{\lower.5ex\hbox{\gtsima}}
\def\ms{$M_{\odot}$ }
\def\msp{$M_{\odot}$}

\slugcomment{Accepted for publication in ApJ Letters}

\begin{document}
\title{Surface pollution of main-sequence stars through encounters with AGB ejecta in $\omega$ Centauri}

\author{Takuji Tsujimoto$^1$, Toshikazu Shigeyama$^2$, and Takuma Suda$^2$}

\altaffiltext{1}{National Astronomical Observatory, Mitaka-shi,
Tokyo 181-8588, Japan; taku.tsujimoto@nao.ac.jp}
\altaffiltext{2}{Research Center for the Early Universe, Graduate School of Science, University of Tokyo, 7-3-1 Hongo, Bunkyo-ku, Tokyo 113-0033, Japan; [shigeyama, suda]@resceu.s.u-tokyo.ac.jp.}

\begin{abstract}
The origin of a double main-sequence (MS) in $\omega$ Centauri is explored. We have shown from theoretical calculations on the stellar evolution that the colors of MS stars are shifted to those of the observed blue MS if the surface layers are polluted by He-rich materials with the mass of $\sim 0.1$ \msp. Stars are supposed to be polluted through numerous encounters with the ejecta descended from massive asymptotic giant-branch (AGB) stars. Two populations of stars with different kinematics exceptionally observed in $\omega$ Cen indicate that kinematically cooler stars are more polluted through encounters with AGB ejecta than kinematically hotter ones because the accretion rate is inversely proportional to the cube of the relative velocity. We propose that both of these factors split the MS in $\omega$ Cen. This theoretical scheme explains why only $\omega$ Cen exhibit a double MS and matches the amount of He necessary to produce the blue MS with that supplied from massive AGB stars. Furthermore, we predict that even if globular clusters (GCs) possess only one generation of stars, the velocity dispersion of stars broaden the MS in the color-magnitude diagram as long as the GCs are massive enough to keep  the AGB ejecta after the burst of star formation. This view explains the broad MS recently found in the GC NGC 2808 which exhibits no scatter in [Fe/H] and thus is likely to consist of a single generation of stars unlike the case of $\omega$ Cen.
\end{abstract}


\section{Introduction}
The color-magnitude diagram (CMD) obtained by recent {\it Hubble Space Telescope} and VLT observations have clearly revealed the existence of a double main-sequence (MS) in the Galactic globular cluster (GC), $\omega$ Centauri \citep{Bedin_04, Sollima_06a}. In addition to the dominant red MS (rMS) population, the blue MS (bMS)  occupies $\sim 20-35\%$ of the MS stars. Surprisingly, spectroscopic observations with GIRAFFE at the VLT have revealed that the rMS has a metallicity of [Fe/H]$\sim -1.6$ with some scatter, whereas  the bMS is more metal-rich: [Fe/H]$\sim -1.3$  \citep{Piotto_05}. Thus, their metallicities do not explain this color difference. A high content of He such as $Y\sim 0.35-0.4$ in the bMS stars is likely to be the only possible origin \citep{Bedin_04, Norris_04}.

Red giants and horizontal-branch (HB) stars in $\omega$ Cen exhibit  a large spread in their metallicities  \citep[e.g.,][]{Norris_95,Suntzeff_96, Sollima_06b}. Especially, the metallicities of red giants spread over $-2\ltsim$[Fe/H]$\ltsim-0.5$, and apparently divide the red giant branch into three distinct sub-populations: the dominant (the fraction $f\sim80\%$) metal-poor ( [Fe/H] $\ltsim$$-1.4$) population; intermediate-metallicity ($-1.4\ltsim$[Fe/H]$\ltsim-0.8$) less populous ($f\sim 15\%$) one  \citep{Norris_96}; extremely metal-rich ([Fe/H]$\sim -0.6$) anomalous one, so called, RGB-a with $f\sim 5\%$ of the entire population \citep{Pancino_00, Ferraro_02}. Furthermore, the difference in stellar metallicities is linked to their kinematics: more metal-rich stars are kinematically cooler than metal-poorer stars \citep{Norris_97}. Such  complex features indicate multiple episodes of star formation in $\omega$ Cen, which is exceptional among GCs because the other Galactic GCs exhibit essentially no metallicity dispersion probably indicating a single generation of stars. The duration of star formation in $\omega$ Cen is inferred  to be a few Gyrs from the observed elemental abundance ratios of red giants \citep{Tsujimoto_03}. In such a scheme of self-enrichment, the bMS stars should have been formed from the gas enriched by the former generations of stars evolved from the rMS. 
This picture naturally provides the condition that the amount ($\Delta Y\sim 0.15$) of the extra He in the bMS stars needs to be less than the amount of He supplied from  massive asymptotic giant-branch (AGB) stars \citep{D'Antona_05} and/or massive stars \citep{Maeder_06} evolved from the rMS population. If  the bMS stars are entirely composed of He-rich gas, this condition has been shown to be hardly satisfied in $\omega$ Cen \citep{Bekki_06} and will be  reviewed in the following section. 
To avoid the short supply of He, \citet{Bekki_06} proposed an alternative scenario in which He is supplied from an external sources in a dwarf galaxy whose nucleus became $\omega$ Cen afterward. This scenario also assumes that the bMS stars are formed from He-rich gases.

In this {\it letter}, we propose a new scenario in which kinematically cooler MS stars enhance their He contents in the surface layers by accreting He enriched matter when they encounter with AGB ejecta. Based on this scenario, we will shed light on the origin of the double MS as well as its uniqueness. As far as we know, $\omega$ Cen is the only object exhibiting a split in the MS. Similarly, NGC 2808 is found to exhibit a broad MS in the CMD, a blue part of which is claimed to be reproduced by a super He-rich population similar to the case of $\omega$ Cen \citep{D'Antona_05,Lee_05}. However, in contrast to $\omega$ Cen, stars of NGC 2808 do not show a metallicity spread \citep{Carretta_04}. Thus,  the double MS should be caused by characteristics unique to  $\omega$ Cen, while the existence of the blue-shifted MS stars should be explained by common features shared with these two GCs.

\section{The amount of He necessary for bMS stars}

Massive AGB stars that have experienced the second dredge-up are raised as a plausible production site of matter with $Y\sim 0.3-0.4$ \citep{D'Antona_05}. Suppose that stars are born exclusively from the ejecta of these AGB stars, then the helium content of the stars will be close to $Y\sim0.4$ and they will populate the bMS. On this occasion, we will check if the total mass of He in AGB ejecta is sufficient for making the observed number of bMS stars in $\omega$ Cen. If we assume that the mass range of stars yielding $Y\sim 0.3-0.4$ is $3.5-8$ \msp,  we obtain the mass fraction $f_{\rm AGB}$=0.08 of AGB stars yielding a high $Y$ in the total stellar mass, adopting the Salpeter initial mass function (IMF).  In terms of supply and demand of He,  the He mass  supplied by such AGB stars is written as $Y\times M_\omega \times f_{\rm rMS} \times f_{\rm AGB}\times f_{\rm ejecta}$ where $f_{\rm ejecta}$ denotes  the mass fraction of AGB ejecta out of the progenitor star (=0.84). On the other hand, the He mass contained in the bMS stars could be written as $Y\times M_\omega \times f_{\rm bMS} $. Thus, these values combined with the observed mass fractions $ f_{\rm bMS}$ and $f_{\rm rMS}$ of the bMS and rMS stars  (=0.25, 0.75) suggest that He is in short supply in $\omega$ Cen, as shown by the relation:  $Y\times M_\omega \times f_{\rm bMS}> Y\times M_\omega \times f_{\rm rMS} \times f_{\rm AGB} \times f_{\rm ejecta}$, where $M_\omega$ denotes the stellar mass of $\omega$ Cen. It has been pointed out that a variation in the IMF can hardly change this magnitude relation \citep{Bekki_06}.

Therefore, it is  implausible that the bMS stars are assumed to be born from pure AGB ejecta.  In the first place, little dilution of AGB ejecta with the interstellar matter before stars are formed, which is required to yield a high-He-content in the bMS stars,  results in their metallicities almost the same as those of the rMS. In other words, the observed abundances of He and Fe in the bMS stars can not be consistently reproduced from this viewpoint. Even if stars were formed from pure AGB ejecta, there would be no 
reason why this happened only in $\omega$ Cen. Besides massive AGB stars,  ($M>10\,M_\odot$) 
with certain rotations are another candidate for yielding a high $Y$ matter  \citep{Maeder_06}. This candidate suffers from a similar shortage of He supply.

\section{Surface Pollution Scenario}
$\omega$ Cen is the only GC in the Galaxy that consists of multiple generations of stars. These generations are also kinematically heterogeneous \citep{Norris_97}. In this chemically and kinematically heterogeneous environment, AGB stars eject their stellar end products including He in the later evolutionary stage, a part of which take the form of planetary nebulae (PNe). The size of the AGB ejecta reaches  $\sim 0.2$ pc which is a typical size of PNe. As a consequence, such He-rich gases will occupy most of the volume of proto-$\omega$ Cen because GCs including $\omega$ Cen are very compact objects whose half radii are a few pc. Thus,  kinematically cool metal-rich stars can accrete He enriched gas  through encounters with these AGB ejecta. Then the accreted matter is mixed inside the surface convective layers and enhances the He contents to make their apparent colors to bluer. In contrast,  metal-poor stars moving faster cannot accrete He enriched gas as much as the metal-rich stars do because the accretion rate is inversely proportional to the cube of the velocity \citep{Bondi_52}.

This occurs exceptionally in a massive GC such as $\omega$ Cen because its deep gravitational potential can retain AGB ejecta.  In this scenario the abundance of He is enhanced only in the outer layer of the metal-rich and kinematically cool MS stars. Therefore, it significantly reduces the amount of He needed to shift stellar colors  to bluer. In order to explain the observed split of the MS that reaches down to $V_{\rm 606}\sim 22-23$ \citep{Bedin_04}, the necessary mass $M_{\rm acc}$ of  matter with $Y=0.4$ accreted to each metal-rich star is found to be $\sim 0.1$ \ms as shown in the next section. This is larger than the typical mass of convective envelope in low mass MS stars. Then the amount of ejecta from massive AGB stars of all generations of stars is found to be $f_{\rm AGB} \times f_{\rm ejecta} \times M_\omega\sim 0.07M_\omega$, which is greater than the mass supposed to be accreted by the bMS stars $f_{\rm bMS}\times M_{\rm acc} \times M_\omega /M_{\rm star}\sim 0.03M_\omega$. Here  $M_{\rm star}$ denotes the average mass of a star (=0.8 \msp), $M_{\rm acc} $ the mass of the AGB ejecta accreted to the star (=0.1 \msp).

There exists supporting evidence for the surface pollution scenario. RR Lyrae stars with their metallicities of [Fe/H]$\sim -1.2$ corresponding to that of the bMS stars are found to have no sign of the He enhancement \citep{Sollima_06b}. It is a natural consequence from the surface pollution scenario because stars increase the masses of their surface convective layers and thus tend to erase the enhancement of He as they evolve from the MS. On the contrary, the observed fact that extreme (very hot) HB stars occupy a fraction of $\sim 30 \%$ \citep{D'Cruz_00} similar to bMS stars in $\omega$ Cen may imply the link between bMS stars and extreme HB stars. This feature is explained in the framework of  fully He-rich stars \citep{Norris_04, Lee_05}.  On the other hand, it has been suggested that the deep mixing in red giants is a promising mechanism for abundance anomalies of red giants in GCs \citep{Kraft_94} and hence it can be related to the extreme HB \citep[e.g.,][]{Sweigart_97, Suda_06}. Unravelling the second parameter problem is crucial to identify which mechanism is responsible for extreme HB stars in $\omega$ Cen.
  
\section{Accretion of AGB ejecta}
In this section, we will estimate the mass accreted by a kinematically cool star in $\omega$ Cen when it passes through He-rich AGB ejecta.  A part of the envelopes of intermediate-mass stars are ionized during the mass-losing process at the post-AGB stage and are visible as PNe. Though the total mass ejected from an AGB star should be a few or several solar masses, observations of Galactic PNe have revealed only 0.2 \ms as a luminous mass on average. That is a so-called missing mass problem in PNe. It has been suggested that most of the missing mass forms molecular clumps. Such a dense form of AGB ejecta is possibly inclined to survive without dispersing into the ISM for a prolonged time. Thus, we assume that most of AGB ejecta take a form of molecular clumps distributed over a region with the size $r_{\rm ejecta}$ corresponding to that of PNe and survive for a period $\tau_\omega$ of star formation in $\omega$ Cen that should be $\sim 2-4$ Gyrs \citep{Tsujimoto_03, Stanford_06}. At the end of the star formation, the remaining gas and AGB ejecta are supposed to be stripped during a passage through the Galactic disk \citep{Tsujimoto_03}. This scheme suggests that the surface metallicity of the bMS stars is converged to  the average value of $\omega$ Cen \citep[$<{\rm [Fe/H]}> \sim -1.4$,][]{Norris_96} because massive AGB ejecta  descended from all populations can pollute the surfaces of stars. That is compatible with no detection of any metallicity spread in the bMS stars \citep{Piotto_05}. 

The total mass $M_{\rm acc}$ accreted to a star is expressed as
\begin{equation}
M_{\rm acc}=dM_{\rm acc}/dt\cdot \tau_{\omega}/t_{\rm enc}\cdot 2r_{\rm ejecta}/v,
\end{equation}
\noindent where $t_{\rm enc}$ denotes the timescale for the encounter between a star and AGB ejecta and $v$ is their relative velocity. Each factor in the right hand side is the mass accretion rate, the total number of encounters per star, and the time for a star to traverse the region of AGB ejecta. First, the accretion rate due to the Bondi accretion is written as
\begin{equation}
dM_{\rm acc}/dt=\pi r_g^2 \rho_{\rm ejecta}v,
\end{equation}
\noindent where $r_g$($=GM_{\rm star}/v^2$) denotes the accretion radius of a star and $\rho_{\rm ejecta}$ is the density of the AGB ejecta averaged over the region with a radius of $r_{\rm ejecta}$. To evaluate the accreted mass, we adopt input values of $M_{\rm star}=0.8$ \msp, $v= 10$ km s$^{-1}$, and  $r_{\rm ejecta}=0.2$ pc together with the mass of AGB ejecta $=4.2$ \ms which is deduced from  the average mass of 3.5-8 \ms AGB stars ($=5$ \msp) and the mass of  a white dwarf  ($=0.8$ \msp). Then, we obtain
\begin{equation}
dM_{\rm acc}/dt \sim 3.3\times10^{14}\, {\rm g/s}.
\end{equation}
Next, the timescale $t_{\rm enc}$  is obtained by
$t_{\rm enc}=(n_{\rm ejecta}\sigma_{\rm ejecta}v)^{-1}$,  where $n_{\rm ejecta}$ is the number density of AGB stars that can  supply He in the proto-$\omega$ Cen whose radius is $r_\omega$, and $\sigma_{\rm ejecta}$ is the cross section of each ejecta given by $\pi r_{\rm ejecta}^2$. Here we assume the constant $n_{\rm ejecta}$ for all period of $\tau_\omega=2$ Gyr. The total number of AGB stars is estimated to be $3.3\times 10^4$ in the proto-$\omega$ Cen with the mass of $2\times10^6$\msp, assuming the Salpeter IMF and the mass range of 3.5-8 \ms relevant to the major He contributor. This number together with $r_\omega=6$ pc which corresponds to the present half radius of $\omega$ Cen yields $n_{\rm ejecta}\sim 1.4\times10^{-54}{\rm cm^{-3}}$. Then, we obtain $t_{\rm enc} \sim  2.2\times 10^3\, {\rm yr} $, which gives $9.2\times10^5$ encounters in total for the period of $\tau_\omega$.
Finally, combined with $2r_{\rm ejecta}/v \sim 1.2\times10^{12} \rm {s}$, we deduce $M_{\rm acc} \sim 3.6\times10^{32} \,{\rm g} \sim 0.18\,M_\odot$. The derived accretion mass is comparable to the mass of the convective envelope of a star with the mass of $\sim$0.4 \msp. This amount of the accreted mass is sufficient to realize a double MS which continues from the MS turn-off down to $V_{\rm 606}\sim22-23$ in the CMD.

\section{Colors of stars polluted by He-rich matter}
\subsection{Isochrone}
\label{isochron}

\begin{figure}[t]
\vspace{0.2cm}
\begin{center}
\includegraphics[width=7cm,clip=true]{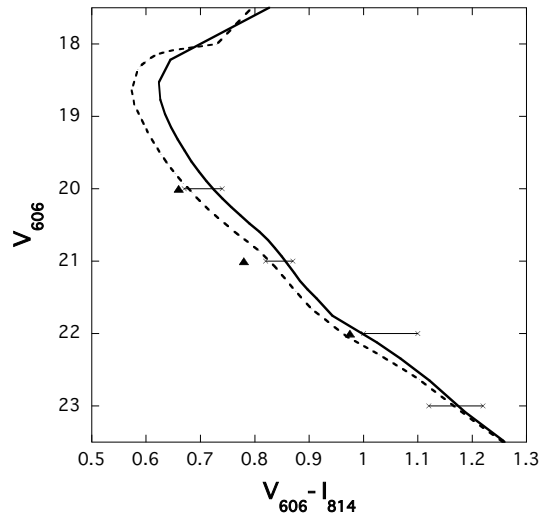}
\end{center}
\vspace{0.3cm}
\caption{Theoretical CMDs of the main sequence of $\omega$ Cen calculated
for the assembly of stars with $Y=0.24$ (solid curve) and of those stars
that accrete the $Y=0.4$ matter by the amount of 0.1\ms (dashed curve).
For reference, the observed points (filled triangles) and ranges
(crosses and bars) are shown at four visual magnitudes of 20, 21, 22,
and 23 mag for bMS and rMS, respectively. These observed data are
extracted from Fig.~1c of \citet{Bedin_04}.}
\end{figure}

We calculate the stellar evolution of low-mass stars with the masses of 0.3 to 0.85 \msp, the surfaces of which are polluted by a $Y=0.4$ matter. The mass accreted to a star is assumed to be 0.1 \msp. For simplicity, we neglect a duration of mass accretion and thus place the $Y$=0.4 matter on the surface layer of a star at the start of calculations. These results are compared with those for stars without a pollution. The stellar evolution program and input physics are the same as in \citet{Suda_04}. The initial abundances are set to be $Y$=0.24 and [Fe/H] = $-1.6$, $-1.3$ for two populations, and  $Y$=0.40 and [Fe/H] = $-1.3$ for the accreted matter. All models are assumed to be $\alpha$-enhanced so that [$\alpha$/Fe]=0.4.
These computations do not include any special treatment for convection. The border of convective zones are determined by the Schwarzschild criterion with mixing length parameter of 1.5.
Isochrones are obtained through the conversion of calculated effective temperatures and luminosities into the color  $V_{606}-I_{814}$ and the magnitude $V_{606}$ of the HST filters according to the recipe of \citet{Holtzman_95}. Extinctions are also calculated from the values given in \citet{Holtzman_95}. Here we adopt the distance modulus ($m-M$)=13.4 and the reddening $E(B-V)$=0.15.  The ages of the two populations are assumed to be 13 Gyr with $\Delta({\rm Age})$=0 since red giants corresponding to both populations exhibit Type II supernova-like abundance patterns suggesting that the age difference should be less than $\sim1$ Gyr and could be ignored.

Figure 1 shows the $V_{606}-I_{814}$ vs.~$V_{606}$ diagram predicted for $Y$=0.24 stars (solid curve) and stars with a 0.1\ms accretion (dashed  curve), together with yardsticks  and filled triangles indicating the observed locations of the rMS and the bMS, respectively \citep{Bedin_04}. The isochrone of stars with the accretion lies close to that of the observed bMS. In our models, the MS split diminishes at around  $V_{606}$ $\sim$ 23, which is compatible with the observation.

\subsection{Double main sequence}

\begin{figure}[t]
%\vspace{-2.5cm}
\begin{center}
\includegraphics[width=7cm,clip=true]{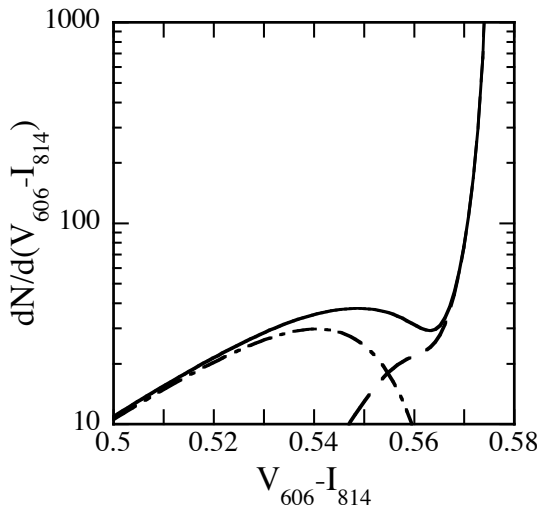}
\end{center}
\vspace{0.5cm}
\caption{The distribution functions  with respect to the color $V_{606}-I_{814}$ of main sequence stars accreting He-rich AGB ejecta. The dashed line shows the distribution function of main sequence stars with [Fe/H]$=-1.6$ and the velocity dispersion of $\sigma_{\rm r}=60$ km/s and the dash-dotted line the distribution function of main sequence stars with [Fe/H]$=-1.3$ and $\sigma_{\rm b}=10$ km/s. The sum of these two populations are shown by the solid line. The number fraction of the blue main sequence stars is assumed to be 35\%. The total number of stars are arbitrary.
}
\end{figure}

The actual color shift of stars due to the surface pollution discussed in the preceding sections should be varied because each star moving at a different speed accretes He-rich gas with a different mass accretion rate. We have derived the distribution function of such stars with respect to the color ($V_{606}-I_{814}$) using the formula
\begin{equation}
\frac{dN}{d\left(V_{606}-I_{814}\right)}=\left|\frac{dN}{dv}\frac{dv}{dM_{\rm acc}}\frac{dM_{\rm acc}}{d\left(V_{606}-I_{814}\right)}\right| ,
\end{equation}
where $dN/dv$ denotes the distribution function of stars with respect to the velocities. Here we assume the isothermal distributions with  constant velocity dispersions $\sigma_{r}$ for the rMS stars  and $\sigma_{b}$ for the bMS stars. The accretion rate due to the Bondi accretion specifies the functional form of $dv/dM_{\rm acc}$, which is proportional to the biquadratic of the velocity. The third factor is derived from the stellar evolution calculations (see \S\ref{isochron}) for stars with the same total mass but  with different masses of the accreted He-rich matter. Since the derived distribution is for stars with the same total mass, this will give the color distribution of stars with the approximately same $V_{606} (=19.3\pm0.2)$. 

Though the colors of the rMS stars are blue-shifted  as well as the bMS stars, the amount of the shift is significantly larger for the bMS. This is because the velocity dispersion $\sigma_{\rm b}$ for stars with [Fe/H]$\sim -1.3$ is smaller than $\sigma_{\rm r}$ for metal-poor stars. A larger velocity dispersion reduces the mass of accreted gas and hence the amount of the color shift due to the He enhancement. For example, the distribution with respect to the color for stars with the mass of $0.77\,M_\odot$ is presented in Figure \ref{colordist}. Here the velocity dispersions are assumed to be $\sigma_{\rm r}=60$ km s$^{-1}$ and $\sigma_{\rm b}=10$ km s$^{-1}$. The distribution has a clear gap between the rMS and bMS stars. Although the value of $\sigma_r$ seems extremely large in comparison with the present velocity dispersion in $\omega$ Cen, an elemental abundance analysis is suggestive of $\sigma_{\rm r}\sim  50$ km/s in the proto-$\omega$ Cen \citep{Tsujimoto_03}. The distribution function of the rMS stars presented in Figure \ref{colordist} has a too sharp peak as compared with the observations \citep{Sollima_06a} because our calculations ignore the scatter in metallicity observed for the rMS stars. This scatter will broaden the rMS to redder parts of the CMD.

Finally, it should be of note that our scenario implies that some GCs with a single generation of stars hold a broad MS rather than a double MS as long as these GCs are massive enough to retain AGB ejecta in their deep gravitational potentials and stars can pass through AGB ejecta many times. 

\section{Conclusions and Discussion}
We propose a new mechanism to produce a double MS in the CMD of $\omega$ Cen, focusing attention on the fact that  high He abundances in the stellar surface layers increase the surface temperatures. On the other hand,  all the other schemes proposed so far assume that stars in the bMS have high He abundances not only in the surface layers but in the entire volumes. However, these schemes confront severe difficulties; (i) the amount of He supplied from AGB stars and/or supernovae evolved from the red MS population would not be sufficient to produce all the bMS stars, (ii) the observed abundances of He and Fe in the bMS stars can not be consistently reproduced, and (iii) there is no reason why this rare phenomenon occurred only in $\omega$ Cen. These critical issues are settled in our scenario as follows. Kinematically cool stars accrete a sufficient amount of AGB ejecta to become the bMS stars. The surface abundance of Fe of these stars converges to the average value of $\omega$ Cen. In contrast, kinematically hot stars accrete too little AGB ejecta to significantly change the colors. Detailed elemental information besides He or Fe for individual MS stars in $\omega$ Cen will give a deeper insight into the origin of  blue MS stars. In fact, a preliminary result on the average C and N abundances of MS stars seems to pose a problematic challenge to our scenario \citep{Piotto_05}.

Generally, a GC undergoes a single burst of star formation because subsequent supernova explosions heat and evaporate the interstellar gas from the GC and thereby stars never accrete He-rich gases. On the other hand, some GCs are expected to be massive enough to retain the interstellar gas \citep{Tsujimoto_03} as well as the AGB ejecta and allow stars to accrete He-rich gas through encounters with the AGB ejecta. Among these GCs, $\omega$ Cen is an exceptional case in the sense that it has multiple stellar populations with different kinematics. Here we have shown that this environment surely  results in a split in MS. The mechanism presented here predicts that other massive GCs with sufficiently deep gravitational potentials to hold AGB ejecta host bluer MS stars that accreted He-rich gas even when no kinematically different population exists. Each star in a GC has a different velocity obeying a certain distribution and thus accretes a different amount of AGB ejecta. NGC 2808 possibly belongs to these massive GCs without multiple kinematic populations.

\acknowledgements

We are grateful to the anonymous referee for the useful comments that helped improve this paper. T. T. acknowledges the Aspen Center for Physics where this work has been stimulated and progressed. T. Shiegayma is grateful to J. E. Norris for informing him on this enigmatic observational results for $\omega$ Cen. T. Suda is grateful to M. Y. Fujmoto for useful comments on accretion model.This work has been partially supported by a grant-in-aid (16540213) of the Japanese Ministry of Education, Science, Culture, and Sports.

\end{document}